\documentclass[a4paper,11pt]{article}
\pdfoutput=1 % if your are submitting a pdflatex (i.e.~if you have
\usepackage{jheppub,amsthm} % for details on the use of the package, please
\usepackage[utf8]{inputenc}

\usepackage{booktabs,multirow, quotes}
\usepackage{chngcntr}
\usepackage[english]{babel}
\usepackage{amsmath,amssymb,graphicx,xcolor,alltt,tikz,subcaption}
\usepackage{framed}

\theoremstyle{remark}

\theoremstyle{theorem}

\makeatletter
\def\@fpheader{\ }
\makeatother

\counterwithout{figure}{section}
\title{Lifshitz critical points meet Zamolodchikov perturbation theory}

\author{Ant\'onio Antunes$^z$}

\affiliation{$^z$Laboratoire de Physique de l'\'Ecole Normale Sup\'erieure, Universit\'e  PSL, CNRS, Sorbonne Universit\'e, Universit\'e  Paris Cit\'e, 24 rue Lhomond, F-75005 Paris, France}

\emailAdd{antonio.antunes@ens.fr}

\abstract{Critical points of classical and quantum lattice models are often described by scale-invariant Lifshitz theories which are anisotropic in the continuum limit, as characterized by a dynamical critical exponent $z\neq1$. This type of critical behavior can in principle be studied by deforming ordinary $z=1$ conformal field theories (CFTs) by relevant vector operators breaking the rotational/Lorentz symmetry. In this short note, we consider a two-dimensional system of coupled minimal model CFTs $\mathcal{M}_{m,m+1}$ which realizes this perspective in a controlled fashion via Zamolodchikov's large $m$ expansion. The model turns out to exhibit interesting properties, including a manifold of interacting Lifshitz fixed points and emergent rotational symmetry in the infrared.  }

\begin{document} 
	
	\maketitle
	
	\flushbottom

\flushbottom

\section{Introduction}
\label{sec:intro}
Critical points of lattice models come in one of two apparently unrelated flavors. In equilibrium statistical physics we have phase transitions driven by thermal fluctuations and described by euclidean quantum field theories (QFTs) in the continuum limit. In quantum many-body physics we have zero-temperature phase transitions driven by quantum fluctuations and described by Lorentzian QFT. Such critical points realize scale invariant physics as guaranteed by the vanishing of the $\beta-$functions, and often have emergent rotational/Lorentz invariant physics, being described by CFTs.\footnote{ \label{foot1}Whether scale+rotational invariant theories are naturally conformal is an interesting question which we will only tangentially address in this note (See for example \cite{Polchinski:1987dy,Dymarsky:2013pqa,Nakayama:2013is,Gimenez-Grau:2023lpz} for useful references). Instead we are interested in the naturality of rotations/Lorentz transformations themselves.} At first glance, this might appear completely obvious for the euclidean case and miraculous for the Lorentzian case. As we zoom out, the  details of the euclidean lattice seem to disappear, while instead, the relation between time and space under a quantum Hamiltonian generating continuous time evolution with a discrete spacial lattice appears anything but symmetric. 

To resolve this apparent dichotomy, we resort to the Wilsonian paradigm, which in this context simply states that:\footnote{Alternatively, rotational/Lorentz invariance can also be achieved if the microscopic Hamiltonian does not overlap with the symmetry breaking operators in the continuum limit.}
\begin{center}
    \textit{Rotation/Lorentz invariance is natural in the RG sense if there are no relevant rotation/Lorentz breaking operators.}
\end{center}
In this paper, we will use the above Wilsonian logic as a starting point. We will begin in the UV with a rotational/Lorentz invariant CFT which contains relevant operators breaking this symmetry: relevant vectors/spin-1 operators. After an RG flow, putative fixed points are expected to break rotational/Lorentz symmetry and will instead be Lifshitz symmetric. We will schematically write
\begin{equation}
    S_{\rm{Lifshitz}}= S_{\rm{CFT}} + g^{\mu}\int d^Dx\,V_\mu(x)\,,  
\end{equation}
where $\Delta_V<D$, which is allowed in unitary CFTs as long as $\Delta_V\geq D-1$. Lifshitz fixed points exhibit translational symmetry and scale invariance, where the scale transformations act as:
\begin{equation}
\label{Lifrescale}
   x^\mu= (\tau,x^i) \to (\lambda^z\tau,\lambda \,x^i)\,,
\end{equation}
where $\tau$ is a privileged direction which could be an arbitrary direction in euclidean signature or time in Lorentzian signature and $i=1,\dots,D-1$. The constant $z$ is known as the dynamical critical exponent and is a key property of the model characterizing its anisotropy in the Euclidean case and its non-relativistic nature in the Lorentzian case. Clearly, for generic values of $z$, Lifshitz scalings are incompatible with rotations/boosts relating the $x_0$ and the $x_i$ directions. Such additional symmetries are only possible for special values of $z$, such as $z=1$ where we have the usual relativistic CFTs, or for $z=2$ where Galilean boosts emerge \cite{NRSon,Baiguera:2023fus,Grinstein:2018xwp,Boisvert:2025hex}. 

In this work, we will focus on $D=2$ systems to take advantage of the existence of a large class of interacting but exactly solvable CFTs to serve as non-trivial UV starting points for the RG flow. This approach is certainly not new and was brilliantly used by Cardy in his study of the chiral Potts model \cite{Cardy:1992tq}. However, as we will review below, this model is rather exceptional due to chirality and integrability, meaning that Cardy's approach does not easily generalize to other cases. Instead, we will make use of the general machinery of conformal perturbation theory to analyze spin-1 deformations which are weakly relevant, admitting a controlled $\epsilon-$like expansion. In particular, inspired by recent work on coupled minimal models \cite{Antunes:2022vtb,Kousvos:2024dlz,Antunes:2024mfb,Antunes:2025huk,Antunes:2025erb} and by Zamolodchikov's study of short RG flows \cite{Zamolodchikov:1987ti}, we will take two copies of a general unitary minimal model $\mathcal{M}_{m,m+1}$ and introduce a relevant vector operator coupling the two copies which is weakly relevant in the large $m$ limit. We will find an infinite family of new Lifshitz fixed points with 
\begin{equation}
    z= 1+ \frac{3}{2\pi m^2}+O(m^{-3})\,,
\end{equation}
which in fact lie in continuous manifolds related by an exactly marginal `nudge' operator which determines the preferred direction of explicit symmetry breaking. Furthermore, it turns out that these fixed points are actually RG unstable: if the system is not fine-tuned it flows to a rotationally invariant CFT and exhibits emergent Lorentz symmetry!

\medskip

The rest of the paper is structured as follows: In Section \ref{sec:LifRev} we review some basic properties of Lifshitz fixed points including Cardy's description of the chiral Potts model. In Section \ref{sec:ConfPert} we introduce the necessary conformal perturbation theory techniques to study deformed conformal field theories. In particular we describe how to generalize the standard analysis to spinning deformations. The main results of the paper are discussed in Section \ref{sec:Zamolodcha}, where we introduce our Lifshitz-Zamolodchikov model: two copies of Virasoro minimal models coupled by a weakly relevant spin-1 operator. We determine the RG diagram of this system, observing manifolds of Lifshitz fixed points and emergent rotation/Lorentz symmetry. We conclude in Section \ref{sec:Conclusions} where we also point out other possible applications of our approach.
\section{Brief review of Lifshitz theory }
\label{sec:LifRev}
In this section we will do a lightning review of some aspects of Lifshitz fixed points. In Section \ref{ssec:basics} we recall some simple physical aspects, in Section \ref{ssec:Examples} we briefly browse through some important examples of Lifshitz theories and in Section \ref{ssec:chiralPotts} we review Cardy's analysis of the chiral Potts model \cite{Cardy:1992tq} which is close in spirit to the Lifshitz-Zamolodchikov model we will introduce in Section \ref{sec:Zamolodcha}.
\subsection{Basic properties}
\label{ssec:basics}
A Lifshitz fixed point is defined by its invariance to rescalings of the form \eqref{Lifrescale}, which connect units of length and energy. Indeed, for a critical system at finite system size $L$, the energy gap to the $i$-th excited state $\Delta E_i$ closes as\footnote{See for example \cite{Wang_2022} for a discussion.
} 
\begin{equation}
    \Delta E_i=\nu \frac{\Delta_i}{L^z}+\dots \,,
\end{equation}
where $\Delta_i$ is the scaling dimension of the associated operator, $\nu$ is a non-universal constant and the ellipsis denote terms that go to zero faster than $L^{-z}$. 
The invariance of the theory under the Lifshitz spacetime symmetry dictates the covariance of the correlation functions of scalar local operators in the continuum limit.\footnote{We will treat all operators as scalars. In general, operators transform in representations of the residual symmetry $SO(D-1)$, but this is trivial in our case of interest $D=2$.} We have
\begin{equation}
   \langle\mathcal{O}(\tau_1,x_1)\mathcal{O}(\tau_2,x_2) \rangle = \frac{1}{\tau_{12}^{2\Delta_\mathcal{O}/z}}\Phi\left( \frac{|x_{12}|^z}{|\tau_{12}|}\right)\,,
\end{equation}
where the scaling function $\Phi$ is a priori theory and operator dependent, and is therefore an interesting observable. It can however be fixed in the limiting cases of small and large argument $u$. Indeed, scaling and translations imply that
\begin{equation}
    \Phi(u) \sim \Phi_0\,, \quad u\to0\,,
\end{equation}
and
\begin{equation}
    \Phi(u) \sim \Phi_\infty u^{-2\Delta_{\mathcal{O}}/z}\,, \quad u\to\infty\,,
\end{equation}
where the ratio $\Phi_0/\Phi_\infty$ is a physical observable. 

Certain special values of $z$ are compatible with an enhanced symmetry algebra which extends dilations and translations. For example for $z=2$, we have the so-called Schr\"{o}dinger algebra, which includes Galilean boosts and a non-relativistic version of special conformal transformations \cite{NRSon}. These additional symetries are sufficient to fix the form of the scaling function $\Phi$. Concretely,
\begin{equation}
     \langle\mathcal{O}(\tau,x)\mathcal{O}(0,0) \rangle = \frac{1}{\tau^{\Delta_\mathcal{O}}} e^{-mx^2/(2 \tau)}\,, \quad(z=2)\,,
\end{equation}
where $m$ is the Galilean invariant mass, which is associated to a central element in the Galilean/Schr\"{o}dinger algebra sometimes referred to as the Bargmann central extension \cite{Bargmann:1954gh}. Remarkably, for more general special values of $z$ of the form $z=2/N$ with $N\in\mathbb
{N}$ and $D=2$, Henkel was able to fix the functional form of the scaling function $\Phi(u)$ in terms of generalized hypergeometric functions $_2F_{N-1}$ \cite{Henkel:1997zz}. We also note that in the case of anisotropy along light-like/holomorphic coordinates in $D=2$, the special `chiral anisotropy' limit $z\to0$, is ensured by general theorems to have the scaling symmetry enhanced to a full local conformal symmetry \cite{Hofman:2011zj}, in analogy to Polchinski's result on the enhancement from ordinary scale to conformal symmetries \cite{Polchinski:1987dy}.

\medskip

Before reviewing some simple Lifshitz invariant models, let us briefly discuss how Lifshitz invariance manifests itself at the level of the stress-tensor of the theory. In a CFT, tracelessness of the conserved symmetric stress-tensor
\begin{equation}
   \delta^{\mu\nu}T_{\mu\nu}=T^\mu_\mu=0 \,, \quad (z=1)\,,
\end{equation}
is a sufficient condition\footnote{The fact that is not a necessary condition is what leads to the non-trivial relation between scale invariance and conformal invariance mentioned in footnote \ref{foot1}.} for the existence of a conserved current $j_\mu$ associated to scale invariance. Indeed, an infinitesimal scale transformation is generated by the conformal Killing vector field $\xi^\mu=x^\mu$\, and then
\begin{equation}
    \partial^\mu j_\mu= \partial^\mu(T_{\mu\nu} x^\nu)=(\partial^\mu T_{\mu\nu}) x^\nu + T_{\mu\nu}\delta^{\mu\nu}=0\,.
\end{equation}
Similarly, an infinitesimal Lifshitz scale transformation \eqref{Lifrescale} is generated by a vector field 
\begin{equation}
    \xi^\mu_z= x^\mu + (z-1)\tau\delta^\mu_\tau\,,
\end{equation}
which then leads to a conserved current $j_\mu^z$ if the stress-tensor satisfies
\begin{equation}
\label{Liftrace}
    z\,T_{\tau \tau}+\delta^{ij}T_{ij}=0\,,
\end{equation}
which we call the Lifshitz-traceless condition.\footnote{See \cite{Arav:2016akx} for a detailed discussion centered on anomalies.} This condition could have also been absorbed by a shift in the metric and will be crucial for computing the dynamical critical exponent $z$ in the models we will study in Section \ref{sec:Zamolodcha}.
\subsection{Example models}
\label{ssec:Examples}
It is straightforward to write down Gaussian models invariant under Lifschitz transformations. Indeed, for a bosonic Lifshitz scalar field $\phi$, we have \cite{Basak:2023otu}
\begin{equation}
    S_{\rm{LS}}= \int d\tau d^dx \, \left(\frac{1}{2} (\partial_\tau \phi)^2 +\frac{v}{2}(\partial_x^z \phi)^2\right)\,,
\end{equation}
which is only local for integer $z$ and $d=D-1$. Here, $v$ is an effective speed of light, which can be renormalized by interactions. Generically, the $z$ exponent can also be renormalized if the interactions do not respect all symmetries. For example, starting with a Schr\"odinger invariant $z=2$ gaussian theory, and adding interactions that break galilean boosts, the exponent $z=2$ is no longer protected and will generally renormalize in the IR. This gaussian action is manifestly Lifshitz invariant for $\phi$ with the scaling dimension
\begin{equation}
    \Delta_\phi=\frac{d-z}{2}\,.
\end{equation}
Remarkably, this Gaussian fixed point actually describes critical points of interesting quantum dimer models, in the case $d=z=2$ \cite{Dimer,Dimer2}. Furthermore, interacting Lifshitz symmetric bosonic systems also describe interesting phase transitions. For example, the transition between the trivial and superfluid phases of the Bose-Hubbard model in $d=2$ is described by an interacting bosonic $z=2$ fixed point \cite{BoseHubbard}. 

Many other Lifshitz critical points occur in fermionic systems. Perhaps the most studied of these is known as `fermions at unitarity' in $d=2$, whose action is given by
\begin{equation}
    S_{\rm{UF}}= \int d\tau d^2x \left( \,\psi^\dagger\partial_\tau \psi + v \,\psi^\dagger\partial_x^2 \psi + \lambda (\psi^\dagger\psi)^2\right)\,,
\end{equation}
where $\psi$ is a spinless fermion. This model is Galilean invariant so has protected dynamical exponent $z=2$ and was discussed for example in \cite{FermiUnitarity,FermiSon,NRSon}.\footnote{See also  \cite{Hellerman:2021qzz,Hellerman:2023myh} for recent studies of this model using the large-charge expansion.} An interesting fermionic model where there is no symmetry protecting $z$ is graphene coupled to an electric field $A^0$, treated as an instantaneous Coulomb interaction, as first introduced by Son in \cite{Son:2007ja}. Introducing $N_f$ flavours of Dirac fermions $\Psi^a, \,a=1,\dots,N_f$, the action reads
\begin{equation}
    S_{\rm{graphene}}=\int d\tau d^2x \left(\bar{\psi}_a \gamma_\tau\partial_\tau\psi_a + v_{F} \bar{\psi}_a \gamma_i\partial_i\psi_a + A^0 \bar{\psi}_a \gamma_\tau\psi_a \right)\,,
\end{equation}
and the dynamical critical exponent can be computed in the large $N_f$ expansion.\footnote{See also \cite{Shpot:2012mw} for a large $N$ analysis of a different Lifshitz model.} To leading non-trivial order, Son found \cite{Son:2007ja}
\begin{equation}
    z=1-\frac{4}{\pi^2N_f}+O\left(\frac{1}{N_f^2}\right) \,.
\end{equation}
An important concept unifying all these models is that perturbation theory is performed around a Gaussian model and, in all except the last example, the dynamical exponent is protected along the RG flow.\footnote{See also \cite{Dhar:2009am,Das:2009fb} for other Lifhsitz models involving fermions and gauge fields. } In the next section, we will review a model with a non-Gaussian UV fixed point where $z$ renormalizes along the flow. We will use this framework as the basis for the remainder of our investigation. 
\subsection{Cardy's approach to the chiral Potts model}
\label{ssec:chiralPotts}
Let us consider the 2d critical 3-state Potts CFT as our UV starting point. This is a non-diagonal unitary minimal model $\mathcal{M}_{5,6}$ with $c=4/5$ and an invertible group symmetry $\mathbb{S}_3$. As a non-diagonal model it posesses spinning Virasoro primary operators, as can be read-off from the following D-type modular invariant partition function  \cite{Cappelli:1986hf}
\begin{equation}
  Z_{\rm{Potts}}(\tau,\bar{\tau})= |\chi_0(\tau)+\chi_3(\tau)|^2+|\chi_\frac{2}{5}(\tau)+\chi_\frac{7}{5}(\tau)|^2+2 |\chi_{\frac{1}{15}}(\tau)|^2+2 |\chi_{\frac{2}{3}}(\tau)|^2 \,,
\end{equation}
where $\tau$ is the usual complex modular parameter on the torus and $\chi_h$ denote Virasoro characters of weight $h$ and $c=4/5$. In particular, there is a relevant spin-1 operator $\Phi_\mu$ with dimension $\Delta_\Phi=9/5$. Let us consider formally adding it to the action as a perturbation of our UV Potts model \cite{Cardy:1992tq}
\begin{equation}
    S=S_{\rm{3-Potts}}+ g^\mu \int d^2x V_\mu(x) = S_{\rm{3-Potts}}+ g^{w} \int \Phi_{w,\,(\frac{7}{5},\frac{2}{5})}+ g^{\bar{w}} \int \Phi_{\bar{w},\,(\frac{2}{5},\frac{7}{5})}\,,
\end{equation}
where in the second equality we introduced (anti-)holomorphic coordinates $w=x+iy$ and $\bar{w}=x-iy$ and specified the left- and right-moving weights $(h,\bar{h})$. A standard analysis suggests that for consistency under RG we should also take into account the operators generated under the OPE of $V$
\begin{equation}
    \Phi_{w,\,(\frac{7}{5},\frac{2}{5})} \times \Phi_{\bar{w},\,(\frac{2}{5},\frac{7}{5})} \sim \epsilon_{(\frac{2}{5},\frac{2}{5})}+ \dots\,\,\,,
\end{equation}
meaning we should consider the more general action
\begin{equation}
  S'  = S_{\rm{3-Potts}}+ g^{w} \int \Phi_{w,\,(\frac{7}{5},\frac{2}{5})}+ g^{\bar{w}} \int \Phi_{\bar{w},\,(\frac{2}{5},\frac{7}{5})}+ g^{\epsilon} \int \epsilon_{(\frac{2}{5},\frac{2}{5})}\,,
\end{equation}
and study the associated RG equations for the three couplings. However, there is a non-invertible topological defect line (TDL) $N$ satisfying a $\mathbb{Z}_3$ Tambara-Yamagami structure \cite{Chang:2018iay} (acting physically as a Kramers-Wannier duality)
\begin{equation}
    N^2=1  +  \eta  +  \eta^2\,,
\end{equation}
where $\eta$ is an invertible TDL generating $\mathbb{Z}_3\subset \mathbb{S}_3$, i.e. $\eta^3=1$. Under this symmetry, the deforming operators transform as
\begin{equation}
    \left[N,\Phi_w \right]= \Phi_w\,, \quad \left[N,\Phi_{\bar{w}} \right]= -\Phi_{\bar{w}}\,,  \quad \left[N,\epsilon \right]= -\epsilon\,,
\end{equation}
meaning that the purely chiral deformation $\Phi_w$ is duality invariant and forms a consistent subspace of the RG equations.\footnote{There is also a TDL $\bar{N}$ under which $\Phi_w$ and $\Phi_{\bar{w}}
$ switch roles, leading to an analogous anti-chiral RG flow.} Indeed, writing down the $\beta$-function equation for the chiral coupling
\begin{equation}
    \beta^w=\frac{1}{5}g^w \,,
\end{equation}
we see that the equation is exact to all orders in perturbation theory as higher order terms in $g^w$ can never transform as spacetime vectors. Therefore, we find the rotationally invariant UV fixed point $g^w=0$ and the anisotropic fixed point, which in this scheme sits at $g^w\to\infty$. This so-called chiral Potts fixed point has a well-known lattice description which satisfies integrable properties \cite{Baxter:1989vv,Watts:1997sm}. Since in this scheme scalar couplings don't run, scalar operators are expected to have the same dimension as in the usual Potts model. To derive the non-trivial dynamical exponent $z$, we note that the perturbed conservation equation for the stress-tensor $T$ 
\begin{equation}
    \partial_{\bar{w}} T_{ww}= \frac{\pi}{2}\left(1-\frac{7}{5}\right) g^w \partial_w \Phi_w\,,
\end{equation}
is exact to all orders in perturbation theory (as can be argued from dimensional analysis), from which we conclude that in the IR fixed point
\begin{equation}
    T_{\bar{w}w}= - \frac{\pi}{2}\left(1-\frac{7}{5}\right) g^w  \Phi_w\,,
\end{equation}
and similarly that
\begin{equation}
    T_{w\bar{w}}= - \frac{\pi}{2}\left(1-\frac{2}{5}\right) g^w  \Phi_w\,.
\end{equation}
The lack of index symmetry of the stress-tensor signals the breaking of rotational invariance, but the modified condition
\begin{equation}
     T_{\bar{w}w}=-\frac{2}{3} T_{w\bar{w}}\,,
\end{equation}
still ensures that under a combined rotation and dilation of the form
\begin{equation}
    (w,\bar{w})\to (\lambda^3w,\lambda^2 \bar{w})\,,
\end{equation}
the system remains invariant. In the standard normalization, we then read-off the dynamical exponent\footnote{Curiously, this is the same dynamical exponent as the KPZ universality class \cite{KPZ}.}
\begin{equation}
    z=\frac{3}{2}\,.
\end{equation}
Before finishing the section, we note that this Lifshitz symmetry is chiral/holomorphic and cannot be related by a real rotation of frame to an ordinary spatial anisotropy of the type we discussed above and will consider in the remainder of this note.\footnote{However, this is precisely the type of symmetry to which the theorem of \cite{Hofman:2011zj} applies in the $z\to0$ limit. See also \cite{Dijkgraaf:1996iy} for more general chiral deformations of 2d CFT.} Alternatively, in Lorentzian signature, this would amount to anisotropy along light-like coordinates, which cannot be boosted to a timelike-spacelike anisotropy.

\section{Conformal perturbation theory with vector deformations}
\label{sec:ConfPert}
In this section we will review the derivation of the one-loop beta function in conformal perturbation theory and identify the necessary modifications to study deformations by spinning operators.

\medskip

Conformal perturbation theory \cite{Zamolodchikov:1987ti} is a convenient formalism to study RG flows starting from a generic UV CFT deformed by an arbitrary relevant operator. One takes the formal action
\begin{equation}
    S_{CPT}= S_{CFT}+ \sum_i\tilde{g}^i \int d^2x\,\mathcal{O}_i (x)=  S_{CFT}+ \sum_i g^i \int \frac{d^2x}{a^{2-\Delta_i}}\,\mathcal{O}_i (x)\,,
\end{equation}
where $g_i$ are dimensionless couplings, $a$ is a UV scale and the $\mathcal{O}_i$ are local operators of dimension $\Delta_i$ which we take to be scalar for now. To determine the running of the $g^i$ and find associated fixed points, one demands invariance of the partition function under combined infinitesimal rescalings $a\to(1+\delta)\,a$ and a shift of the coupling constants. A weak coupling expansion yields
\begin{equation}
    Z_{\rm{CPT}} \!= \!Z_{\rm{CFT}} \Bigg(1\!-\!\sum_i\!g^i \!\int \!\! \frac{d^2x}{a^{2-\Delta_i}}\langle\mathcal{O}_i (x)\rangle \!+\! \frac{1}{2}\!\sum_{i,j} \!\frac{g^ig^j}{a^{4-\Delta_i-\Delta_j} 
    } \int_{|x-x'|>a} \!\!\!\!\!\!\!\!\!\!\!\!\!\!\!d^2x\,d^2x'\langle\mathcal{O}_i (x)\mathcal{O}_j (x')\rangle\!\Bigg)\,,
\end{equation}
where we dropped cubic and higher-order terms and used the scale $a$ as a position space cutoff to regularize the collision of perturbing operators. When we rescale $a$, the change through the explicit $a$ dependence is immediately absorbed by shifting
\begin{equation}
    g^i\to g^i+\delta\,(2-\Delta_i)  g^i\,.
\end{equation}
Instead, the variation in the regulated integration region is captured by the infinitesimal annulus $a<|x-x'|<a(1+\delta)$ and can be computed using the operator product expansion (OPE)
\begin{align}
\label{OPEbeta}
   &\sum_{i,j} \!\frac{g^ig^j}{a^{4-\Delta_i-\Delta_j} 
    } \int_{a<|x-x'|<a(1+\delta)} \!\!\!\!\!\!\!\!\!\!\!\!\!\!\!\!\!\!\!\!\!\!\!\!\!\!\!\!\!\!\!\!\!\!\!\!d^2x\,d^2x'\langle\mathcal{O}_i (x)\mathcal{O}_j (x')\rangle= \!\sum_{i,j,k} \!\frac{g^ig^j C_{ij}^k}{a^{4-\Delta_i-\Delta_j} 
    } \int_{a<|x-x'|<a(1+\delta)} \!\!\!\!\!\!\!\!\!\!\!\!\!\!\!\!\!\!\!\!\!\!\!\!\!\!\!\!\!\!\!\!\!\!\!\!d^2x\,d^2x'  a^{\Delta_k-\Delta_i-\Delta_j}\langle\mathcal{O}_k (x')\rangle=\nonumber\\
    &= 2\pi\delta\sum_{i,j,k} g^ig^j C_{ij}^k\int \!\! \frac{d^2x'}{a^{2-\Delta_k}}\langle\mathcal{O}_k (x')\rangle \,,
\end{align}
where $C_{ij}^k$ is the OPE coefficient and use the fact that the operators were scalars to replace $|x-x'|\to a$ inside the infinitesimal annulus of area $2\pi  a^2 \delta$.
Performing a second shift in $g^i$ to capture this variation allows us to read-off the beta-function to second order in the coupling
\begin{equation}
\label{oneloopbeta}
    \beta^i(g)=(2-\Delta_i)g^i- \pi \sum_{j,k}C_{jk}^i g^j g^k+O(g^3)\,.
\end{equation}

\medskip

To generalize this analysis to the spin-1 deformations we are interested in, we simply need to adjust the use of the OPE in the first equality in \eqref{OPEbeta}. For operators with general quantum numbers $(h,\bar{h})$, holomorphic factorization dictates the structure
\begin{equation} 
    \mathcal{O}_i(w,\bar{w})\times \mathcal{O}_j(w',\bar{w}')\subset \frac{C_{ij}^k \mathcal{O}_k(w',\bar{w}')}{(w-w')^{h_i+h_j-h_k}(\bar{w}-\bar{w}')^{\bar{h}_i+\bar{h}_j-\bar{h}_k}}\,,
\end{equation}
which leads to a non-vanishing angular integral over the annulus in \eqref{OPEbeta} if and only if
\begin{equation}
    h_i+h_j-h_k=\bar{h}_i+\bar{h}_j-\bar{h}_k\,,
\end{equation}
which amounts to the `conservation of spin' condition
\begin{equation}
    s_i+s_j=s_k\,.
\end{equation}
Therefore, in the one-loop beta function equations \eqref{oneloopbeta}, we can also let the indices $i$ run over the two components $s=\pm1$ of the spin one operators, or more generally the components of any spinning operator, as long as we include the spin constraint. We can therefore write
\begin{equation}
\label{oneloopbetaspin}
    \beta^i(g)=(2-\Delta_i)g^i- \pi \sum_{j,k} \delta_{s_j+s_k,s_i}C_{jk}^i g^j g^k+O(g^3)\,.
\end{equation}
One could have anticipated this result by noting that the couplings to spinning operators $g^\mu$ act as spurions for the $U(1)$ spacetime rotational symmetry. Therefore in the beta-function equations \eqref{oneloopbetaspin} we can have terms such as
\begin{equation}
    \beta^w \subset g^w g^\epsilon\,, \quad   \beta^{\bar{w}} \subset g^{\bar{w}}g^\epsilon \,, \quad \beta^\epsilon \subset g^w g^{\bar{w}}\,,
\end{equation}
but not
\begin{equation}
\label{notallowed}
    \beta^w \not \subset g^{\bar{w
    }} g^\epsilon\,, \quad \beta^w \not \subset g^wg^w
\end{equation}
as they do not respect the custodial $U(1)$ symmetry.\footnote{See also \cite{Iengo:2009ix,Nakayama:2013ssa} for more extensive discussions on renormalization in Lorentz breaking field theories.} 
\section{The Lifshitz-Zamolodchikov model}
\label{sec:Zamolodcha}
In this section we will introduce the Lifshitz-Zamolodchikov model obtained as a weakly relevant spin-1 perturbation which couples two minimal models. We briefly recall Zamolodchikov's analysis of the rotational invariant RG flows between neighbouring minimal models in Section \ref{ssec:oneM} and study the anisotropic coupled model in Section \ref{ssec:twoMs}.
\subsection{Zamolodchikov's perturbation theory}
\label{ssec:oneM}
Conformal perturbation can only be used systematically if the deviation to marginality can be made parametrically small \cite{Komargodski:2016auf}. While the Wilson-Fisher expansion in $\epsilon=4-D$ \cite{PhysRevLett.28.240} is the standard way to achieve this, here we want to take advantage of interacting exactly solvable models and remain in $D=2$. A way to introduce a small parameter while fixing the spacetime dimension $D=2$ was introduced by Zamolodchikov in \cite{Zamolodchikov:1987ti}. One considers the diagonal Virasoro minimal models $\mathcal{M}_{m,m+1}$ in the highly multi-critical regime $m\to \infty$ \cite{Behan:2017rca,Ribault:2018jdv}. In this limit, we have the central charge and Kac weights 
\begin{equation}
\label{chlimits}
    c\approx1-\frac{6}{m^2}\,, \qquad h_{r,s}\approx \frac{(r-s)^2}{4}+ \frac{r^2-s^2}{4m}\,.
\end{equation}
One notices that the scalar operator $\phi_{(1,3)}$ is weakly relevant with $\Delta_{1,3}\approx2-\frac{4}{m}$, and might therefore lead to a short flow under perturbative control. Indeed, since $\phi_{(1,3)}$ satisfies the OPE
\begin{equation}
\phi_{(1,3)}\times\phi_{(1,3)}\sim 1+   \phi_{(1,3)} + \textrm{irrelevant ops}\,,
\end{equation}
there is a consistent RG flow obtained by deforming the minimal model $\mathcal{M}_{m,m+1}$ by a single coupling $g_{(1,3)}$ with the one-loop beta-function
\begin{equation}
    \beta_{(1,3)}=\frac{4}{m}g_{(1,3)}- \pi C_{(1,3),(1,3)}^{(1,3)} g_{(1,3)}^2\,,
\end{equation}
where the OPE coefficient $C_{(1,3),(1,3)}^{(1,3)}\approx4/\sqrt{3}$ is of order 1 in the large $m$ limit \cite{Dotsenko:1984ad,Dotsenko:1984nm}.
This leads to a weakly coupled IR fixed point $g_{(1,3)}^*=\sqrt{3}/(\pi m)$, describing the minimal model $\mathcal{M}_{m-1,m}$. We would like to find a suitable modification of this setup that allows us to deform by a weakly relevant spin 1 operator. Since these models are diagonal, relevant spin 1 operators must be descendants, i.e. total derivatives that do not modify the action. An alternative would be to search for spin 1 primaries in the non-diagonal D-series minimal models \cite{Ribault:2019qrz}, but the marginal spin 1 weights $(h,\bar{h})=(1/2,3/2)$ are not an accumulation point of the Kac table \eqref{chlimits}.

\medskip

\subsection{Coupling minimal models}
\label{ssec:twoMs}
A simple workaround to this obstruction is to consider $N$-fold tensor products of diagonal minimal models \cite{Antunes:2022vtb,Antunes:2024mfb,Antunes:2025huk,Antunes:2025erb}, since additivity allows us to obtain a bigger set of chiral weights under the diagonal Virasoro symmetry generated by $T^{(1)}+T^{(2)}$. Indeed, even in the simplest case of $N=2$ copies/replicas that we will consider for the rest of this note, we can build the operator
\begin{equation}
    V_\mu= \phi_{(1,2)}^{(1)} \partial_\mu \phi_{(1,2)}^{(2)}- \phi_{(1,2)}^{(2)} \partial_\mu \phi_{(1,2)}^{(1)}\,,
\end{equation}
where the superscript is a copy/replica label. This operator is a global Virasoro primary under the diagonal conformal symmetry, has spin 1 and dimension $\Delta_V \approx2-3/m$ and is therefore weakly relevant. We would then like to consider the formal action
\begin{equation}
    S= \sum_{i=1}^2S_m^{(i)} + g^\mu \int d^2x \, V_\mu(x)\,.
\end{equation}
Unfortunately, Zamolodchikov's argument above that led to a consistent RG flow with a single coupling does not hold in this case since as we saw in \eqref{notallowed} this is kinematically forbidden for spin-1 couplings. Furthermore, it is also dynamically forbidden as $V \times V \not \subset V\,,$
which is a consequence of the OPE $\phi_{(1,2)}\times\phi_{(1,2)} \sim 1+ \phi_{(1,3)}$. Fortunately, this obstruction points to the solution: we simply need to include an additional coupling to $\phi_{1,3}$ to consistently truncate the beta-function equations. We dub the resulting theory the Lifshitz-Zamolodchikov model and its action reads
\begin{equation}
\label{LifZam}
    S_{\rm{LZ}}= \sum_{i=1}^2S_m^{(i)} + g^\mu \int d^2x (\phi_{(1,2)}^{(1)} \partial_\mu \phi_{(1,2)}^{(2)}- \phi_{(1,2)}^{(2)} \partial_\mu \phi_{(1,2)}^{(1)})\,+ \frac{g^{\epsilon}}{\sqrt{2}} \int d^2x \,(\phi_{(1,3)}^{(1)}+\phi_{(1,3)}^{(2)})\,,
\end{equation}
which preserves a diagonal spin-flip $\mathbb{Z}_2$ along with an additional permutation $\mathbb{Z}_2$ (which is really a composition of a permutation and a spin flip in a single copy) for odd $m$. We also introduced the shorthand notation $g^\epsilon$ for the $g^{(1,3)}$ coupling. Using the spinning beta-function equations \eqref{oneloopbetaspin}, along with the large $m$ OPE coefficient $C_{(1,2),(1,2)}^{(1,3)}\approx\sqrt{3}/2$ and making sure to work with unit-normalized two point-functions, we find 
\begin{align}
\label{betacoupled}
    \beta_z &= \frac{3}{m}g_z -\sqrt{6} \pi g_\epsilon g_z\,, \nonumber\\
    \beta_{\bar{z}} &= \frac{3}{m}g_{\bar{z}} -\sqrt{6} \pi g_\epsilon g_{\bar{z}}\,, \\
    \beta_\epsilon&= \frac{4}{m}g_\epsilon - 2\pi \sqrt{\frac{2}{3}}g_\epsilon^2 - \sqrt{6} \pi g_z g_{\bar{z}}\,, \nonumber
\end{align}
which leads to three types of fixed points:
\begin{itemize}
    \item The UV fixed point $\mathcal{M}_{m,m+1}\otimes\mathcal{M}_{m,m+1}$ at $g^z=g^{\bar{z}}=g^\epsilon=0$.
    \item A purely scalar deformed fixed point at $g^z=g^{\bar{z}}=0$ and $g^\epsilon=\sqrt{6}/(m\pi)$, which is simply $\mathcal{M}_{m-1,m} \otimes \mathcal{M}_{m-1,m}$, i.e. two decoupled copies of the Zamolodchikov flow.
    \item A circle of non-trivial fixed points with $g^\epsilon= \sqrt{3}/(\sqrt{2}m\pi)$ and $g^zg^{\bar{z}}=1/(m\pi)^2$.
\end{itemize}
The fact that there is a circle of rotational symmetry breaking fixed points given that a single such fixed point exists is unsurprising: Due to UV rotational symmetry, any preferred direction introduced by the spin-1 coupling is equivalent. However it is important to note that, unlike the model discussed in Section \ref{ssec:chiralPotts}, the circle of fixed points $g^zg^{\bar{z}}=(g^x)^2+(g^y)^2=1/(m\pi)^2$ is only compatible with non-chiral anisotropy: Both $g^z$ and $g^{\bar{z}}$ must be non-vanishing at fixed point. 
What might be surprising is that a presumably anisotropic fixed point can exist in an isotropic RG scheme, like the one we have used. The solution to this apparent paradox is simple: There are actually couplings that we have omitted which do not have a vanishing beta-function: the couplings to the stress-tensor operator, i.e. the metric. Indeed, letting 
\begin{equation}
    S_{\rm{LZ}} \to S_{\rm{LZ}}+ g^{\mu \nu} \int d^2x\,T_{\mu \nu}(x)\,,
\end{equation}
 and using our spinning beta-function equations \eqref{oneloopbetaspin}, we find the beta-function for the metric
 \begin{equation}
     \beta^{z z}= -\pi \frac{3}{2}g^z g^z\,,
 \end{equation}
and similarly for $\beta^{\bar{z} \bar{z}}$. Clearly, this beta-function does not vanish at the fixed-point above. However, adding the stress-tensor to the action simply amounts to a shift in the metric, which should be such that we have Lifshitz invariance rather than usual conformal invariance, following the discussion of Section \ref{ssec:basics}. To see this, it is enough to recall that in the isotropic RG scheme \cite{Jack:1990eb}
\begin{equation}
    T^\mu_\mu=\sum_i\beta^i \mathcal{O}_i\,,
\end{equation}
and that we have solved the beta function equations \eqref{betacoupled} to order $1/m^3$. Therefore we can write
\begin{equation}
    T^\mu_\mu - \beta^{zz}T_{zz}- \beta^{\bar{z}\bar{z}}T_{\bar{z}\bar{z}}=O(1/m^3)\,,
\end{equation}
which we can reinterpret as a modified Lifshitz-traceless condition as in equation \eqref{Liftrace}. Taking the spinning deformation to be along the $x$ direction, which can be done without loss of generality, we find 
\begin{equation}
    (T_{xx}+T_{yy})+ \frac{3}{2\pi m^2} T_{xx}=0\,,
\end{equation}
where we plugged in the fixed point coupling. This allows us to determine the dynamical critical exponent
\begin{equation}
    z= 1+ \frac{3}{2\pi m^2} +O(1/m^3)\,.
\end{equation}
It is clear from the calculation above that while the couplings and anomalous dimensions can only be trusted to order $1/m$, the leading contribution to $z$ is at order $1/m^2$ and this term will not get modified at higher order in the perturbative expansion.\footnote{See also \cite{Korovin:2013bua} for a closely related computation in the case of classically marginal spin-1 deformations.} This is entirely analogous to the $c$-theorem sum rules \cite{Zamolodchikov:1986gt,Cardy:1988tj} which reliably  predict an order $1/m^3$ modification of the central charge from an order $1/m$ calculation.

\medskip

Having understood how to make sense of the non-trivial fixed point and its Lifshitz structure, we can study a few more simple observables. In particular, we can compute the close to marginal eigenoperators and their anomalous dimensions by diagonalizing the matrix $\partial \beta_i/\partial g_j$ whose eigenvalues are $\gamma_i=2-\Delta_i^{*}$ . 
We find 
\begin{equation}
\label{eigenval}
    \gamma_1=0\,, \quad\gamma_2= - \frac{2 \sqrt{3}}{m}\,,\quad \gamma_3= +
    \frac{2 \sqrt{3}}{m}\,,
\end{equation}
corresponding to the eigenoperators
\begin{equation}
\label{eigenop}
    \mathcal{O}_1= -g^zV_z + g^{\bar{z}}V_{\bar{z}} \,, \quad \mathcal{O}_2= \sqrt{\frac{g^z}{2 g^{\bar{z}}}} V_z + \sqrt{\frac{g^{\bar{z}}}{2 g^{z}}} V_{\bar{z}} + \epsilon \,, \quad \mathcal{O}_3= \sqrt{\frac{g^z}{2 g^{\bar{z}}}} V_z + \sqrt{\frac{g^{\bar{z}}}{2 g^{z}}} V_{\bar{z}} - \epsilon \,.
\end{equation}
The anomalous dimensions \eqref{eigenval} clearly manifest the broken rotation symmetry, as the two components of the spin 1 deformation which had the same scaling dimension have now split and in fact are mixed with the scalar operator $\epsilon$ as can be seen in \eqref{eigenop}. We also identify a one-loop marginal operator
\begin{equation}
    \mathcal{O}_1= V_\theta\,,
\end{equation}
where $\theta$ is the usual angular coordinate  on the plane such that $z=re^{i\theta}$ and $\bar{z}=re^{-i\theta}$. Indeed, this operator simply rotates the direction along which isotropy is broken, generating a full circle of identical Lifshitz fixed points, which are related by an axis rotation. In the IR fixed point, this simply relabels different components of tensor operators in the UV. This is fully analogous to the exactly marginal tilt operators in global symmetry breaking conformal defects, which generate a defect conformal manifold with identical CFT data but with a rotated basis of operators \cite{Padayasi:2021sik,Drukker:2022pxk,Gabai:2025zcs,Girault:2025kzt,Drukker:2025dfm}. For this reason, we will refer to $\mathcal{O}_1$ as the `nudge' operator and we expect it to be marginal non-perturbatively due to the broken conservation equation for the current associated to rotation symmetry. Crucially, the nudge operator is not null\footnote{Instead, there is a null operator which is a linear combination of $V_\theta$ and $T_{[\mu\nu]}$ (the antissymetric part of the stress tensor, which is null by itself at the UV fixed point). We thank Marco Meineri for a discussion on this point.} as it has non-vanishing correlators in the IR. We should think of it as acting on the coordinates $x^\mu$, which still transform non-trivially under rotations, in the same way that bulk operators transform non-trivially under the full global symmetry $G$ even though the defect breaks the symmetry $H\subset G$.\footnote{See also \cite{Kravchuk:2025evf} for defects breaking transverse rotation symmetry.}

\medskip

\begin{figure}[ht]
\centering
\includegraphics[scale=0.27]{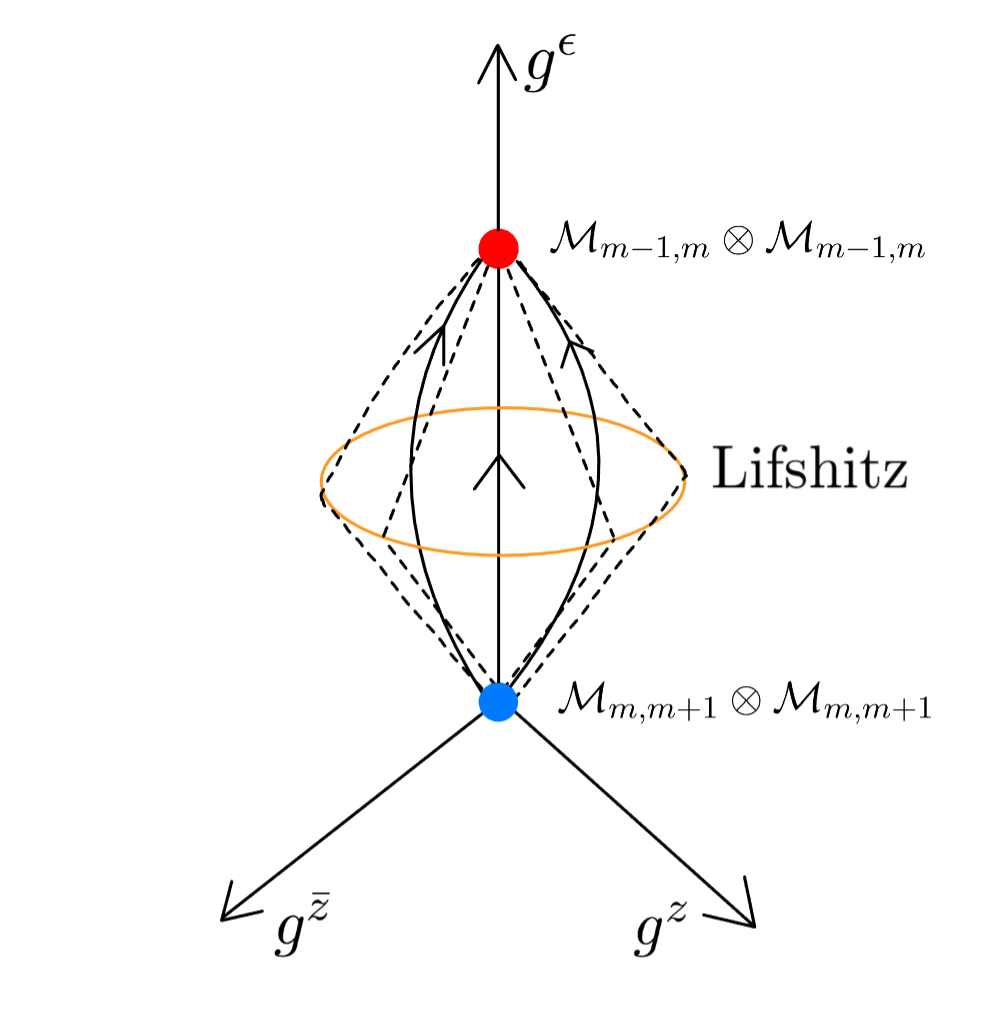}
\caption{RG flow diagram for the Lifshitz-Zamolodchikov model \eqref{LifZam}. The UV fixed point is denoted in blue and the rotationally invariant IR fixed point is denoted in red. Generic flows connect these two points and are denoted by the black arrows. The circle of RG unstable Lifshitz fixed points is denoted in orange and the fine-tuned flows which hit these points are denoted by the dashed lines. }
\label{fig:Lif}
\end{figure}

Before concluding, let us comment on the structure of the underlying RG flow. As is clear from \eqref{eigenval}, the Lifshitz fixed point is RG unstable, as there is still a relevant operator. Instead, the stable fixed point is the rotational invariant CFT $\mathcal{M}_{m-1,m} \otimes \mathcal{M}_{m-1,m}$, which has RG eigenvalues
\begin{equation}
       \gamma_1=-\frac{4}{m}\,, \quad\gamma_2= - \frac{3}{m}\,,\quad \gamma_3=- \frac{3}{m}\,,
\end{equation}
associated to the same eigenbasis of the UV theory, i.e.
\begin{equation}
       \mathcal{O}_1= \epsilon \,, \quad \mathcal{O}_2=  V_z  \,, \quad \mathcal{O}_3= V_{\bar{z}}  \,.
\end{equation}
This means that if we do not fine-tune to hit the Lifshitz theory, our broken rotational symmetry will reemerge in the IR as the system once again flows to decoupled minimal models. This is summarized in Figure \ref{fig:Lif}.

In particular, the RG flow satisfies the $c-$theorem, even though it is not guaranteed by the original proof which assumes rotational invariance \cite{Zamolodchikov:1986gt,Cardy:1988tj}. More generally, it is interesting to ask whether the $c-$theorem always holds even when rotational symmetry is broken at intermediate scales,\footnote{We thank Jo\~ao Penedones from whom we first heard about this question.} but the results of \cite{Swingle:2013zla} seem to suggest otherwise. 
\section{Conclusions and Outlook}
\label{sec:Conclusions}
In this note, we studied weakly relevant spin-1 perturbations that describe RG flows between CFTs and weakly anisotropic Lifshitz fixed points with $|z-1|\ll1$. We introduced a simple system of $N=2$ coupled minimal models $\mathcal{M}_{m,m+1}\otimes\mathcal{M}_{m,m+1}$ where perturbation theory becomes reliable through the introduction of a small parameter $1/m$ and were able to compute $z$ to first non-trivial order in the associated expansion. We also observed the existence of a one-loop marginal nudge operator,\footnote{In the future, it would be interesting to understand the marginality of the nudge operator more deeply and non-perturbatively, perhaps leading to sum-rules similar to those of \cite{Gabai:2025zcs,Girault:2025kzt}.} which rotates the frame of anisotropy and the emergence of rotational symmetry in the infrared. 

It would be interesting to study the more general case of $N$ coupled minimal models, and to study these systems non-perturbatively directly on the lattice. It is straightforward to write down a Hamiltonian for statistical 2d Ising models coupled via a spin-1 operator\footnote{In fact this model is closely related to the axial next-to-nearest-neighbor Ising model (ANNNI) \cite{SELKE1988213}, which was studied in the continuum via bosonization in  \cite{DAllen_2001}. We thank 
Philippe Lecheminant for bringing this to our attention.}
\begin{equation}
    H=  J\sum_{<i,j>}\sigma_i\sigma_j + J\sum_{<i,j>}\tau_i\tau_j + K \sum_i \sigma_i\tau_{i+\hat{n}}\,,
\end{equation}
where $\sigma$ and $\tau$ are two sets of classical Ising variables, $<i,j>$ denotes a pair of nearest-neighbour lattice sites and $\hat{n}$ is a unit vector which selects the privileged direction of anisotropy. However, the large $m$ expansion and the topology of the phase diagram in Figure \ref{fig:Lif} suggest that it might be more prudent to work with coupled tricritical Ising models, perhaps in the Blume-Capel formalism \cite{BlumeCapel}.

Introducing a weakly relevant spin-1 deformation can also be done in more general settings where traditional perturbation theory is possible. Consider, for example the most general Wilson-Fisher model \cite{PhysRevLett.28.240} for $N$ coupled scalars described by the action
\begin{equation}
    S_{\rm{WF}}=\int d^{4-\epsilon}x\,  \left( \frac{1}{2}  (\partial \phi_i)^2  + g_{ijkl} \phi_i\phi_j\phi_k\phi_l\right)\,,
\end{equation}
whose general fixed points have been systematically studied in \cite{Osborn:2017ucf,Rychkov:2018vya,Osborn:2020cnf,Hogervorst:2020gtc,Rong:2023xhz}. If we allow for spacetime rotation symmetry breaking operators, and some amount of global symmetry breaking, we can write down the more general Lifshitz-Wilson-Fisher model
\begin{equation}
    S_{\rm{LWF}}=\int d^{4-\epsilon}x\,  \left( \frac{1}{2}  (\partial \phi_i)^2  + g_{ijkl} \phi_i\phi_j\phi_k\phi_l+ g_{ijk}^\mu(\partial_\mu\phi_i)\phi_j\phi_k\right)\,,
\end{equation}
where only the primary piece of the spin-1 operator can contribute to the action. Classifying Lifshitz fixed points of this model, and identifying potentially relevant phase transitions described by them would be a worthwhile endeavor.

Finally, we would like to mention a more specific application in the realm of $D=2$ physics. While the Coleman-Mermin-Wagner theorem generally forbids long range order for continuous symmetries in $D=2$ euclidean dimensions \cite{MerminWagner,Coleman:1973ci}, it does not strictly apply to quantum $1+1$ dimensional models. In particular, \cite{Nahum:2025bdg} proposed that the $U(1)$ shift symmetry of a compact boson $\varphi$ can be ordered by coupling the field to a critical Ising spin via a Berry phase-like term
\begin{equation}
    S=S_{\varphi}+S_{\rm{Ising}}+ i g\int dtdx \,\sigma\partial_t\varphi\,.
\end{equation}
However, this coupling is strongly relevant ($\Delta=1.125$) and is not under perturbative control. Whether the formalism developed in this paper can be used to turn the insight of \cite{Nahum:2025bdg} into a perturbatively robust framework is currently under investigation.

\paragraph{Acknowledgements}
AA is especially thankful to Jo\~ao Penedones for encouraging the study of Lifshitz fixed points in this context, to Gabriel Cuomo for illuminating discussions and for providing many useful references, to Adam Nahum for patient discussions on related work and to Andreia Gon\c{c}alves for continued inspiration. We further thank Apratim Kaviraj and Miguel Paulos for comments on the draft and Connor Behan, Xiangyu Cao, Subham Dutta-Chowdhury, Vasco Gon\c{c}alves, Johan Henriksson, Apratim Kaviraj, Stefanos Kousvos, Petr Kravchuk, Philippe Lecheminant, John McGreevy, Marco Meineri, Sridip Pal, Miguel Paulos, Alessandro Podo, Giovanni Rizi, Junchen Rong, Slava Rychkov, Adar Sharon, Dam Son and Balt van Rees for useful conversations. AA acknowledges hospitality and funding from Perimeter Institute during the final stages of writing of the draft. AA is funded
by the European Union (ERC, FUNBOOTS, project number 101043588). Views and opinions
 expressed are however those of the author(s) only and do not necessarily reflect those of
 the European Union or the European Research Council Executive Agency. Neither the
 European Union nor the granting authority can be held responsible for them.

\bibliography{main.bib}

@article{KPZ,
author = {CORWIN, IVAN},
title = {THE KARDAR–PARISI–ZHANG EQUATION AND UNIVERSALITY CLASS},
journal = {Random Matrices: Theory and Applications},
volume = {01},
number = {01},
pages = {1130001},
year = {2012},
doi = {10.1142/S2010326311300014}
}

@article{Cardy:1992tq,
    author = "Cardy, John L.",
    title = "{Critical exponents of the chiral Potts model from conformal field theory}",
    eprint = "hep-th/9210002",
    archivePrefix = "arXiv",
    reportNumber = "IN-92003, UCSBTH-92-37",
    doi = "10.1016/0550-3213(93)90353-Q",
    journal = "Nucl. Phys. B",
    volume = "389",
    pages = "577--586",
    year = "1993"
}

@article{Iengo:2009ix,
    author = "Iengo, Roberto and Russo, Jorge G. and Serone, Marco",
    title = "{Renormalization group in Lifshitz-type theories}",
    eprint = "0906.3477",
    archivePrefix = "arXiv",
    primaryClass = "hep-th",
    reportNumber = "SISSA-32-2009-EP",
    doi = "10.1088/1126-6708/2009/11/020",
    journal = "JHEP",
    volume = "11",
    pages = "020",
    year = "2009"
}

@article{Watts:1997sm,
    author = "Watts, G. M. T.",
    title = "{Conserved charges in the chiral 3 state Potts model}",
    eprint = "hep-th/9708167",
    archivePrefix = "arXiv",
    reportNumber = "KCL-MTH-97-51",
    doi = "10.1088/0305-4470/31/25/010",
    journal = "J. Phys. A",
    volume = "31",
    pages = "5599--5607",
    year = "1998"
}

@article{Basak:2023otu,
    author = "Basak, Jaydeep Kumar and Chakraborty, Adrita and Chu, Chong-Sun and Giataganas, Dimitrios and Parihar, Himanshu",
    title = "{Massless Lifshitz field theory for arbitrary z}",
    eprint = "2312.16284",
    archivePrefix = "arXiv",
    primaryClass = "hep-th",
    doi = "10.1007/JHEP05(2024)284",
    journal = "JHEP",
    volume = "05",
    pages = "284",
    year = "2024"
}

@article{Swingle:2013zla,
    author = "Swingle, Brian",
    title = "{Entanglement does not generally decrease under renormalization}",
    eprint = "1307.8117",
    archivePrefix = "arXiv",
    primaryClass = "cond-mat.stat-mech",
    doi = "10.1088/1742-5468/2014/10/P10041",
    journal = "J. Stat. Mech.",
    volume = "1410",
    number = "10",
    pages = "P10041",
    year = "2014"
}

@article{Shpot:2012mw,
    author = "Shpot, M. A. and Pis'mak, Yu. M.",
    title = "{Lifshitz-point correlation length exponents from the large-n expansion}",
    eprint = "1202.2464",
    archivePrefix = "arXiv",
    primaryClass = "hep-th",
    doi = "10.1016/j.nuclphysb.2012.04.011",
    journal = "Nucl. Phys. B",
    volume = "862",
    pages = "75--106",
    year = "2012"
    }

@article{BlumeCapel,
  title = {Finite-size scaling study of the two-dimensional Blume-Capel model},
  author = {Beale, Paul D.},
  journal = {Phys. Rev. B},
  volume = {33},
  issue = {3},
  pages = {1717--1720},
  numpages = {0},
  year = {1986},
  month = {Feb},
  publisher = {American Physical Society},
  doi = {10.1103/PhysRevB.33.1717},
  url = {https://link.aps.org/doi/10.1103/PhysRevB.33.1717}
}

@article{Padayasi:2021sik,
    author = "Padayasi, Jaychandran and Krishnan, Abijith and Metlitski, Max A. and Gruzberg, Ilya A. and Meineri, Marco",
    title = "{The extraordinary boundary transition in the 3d O(N) model via conformal bootstrap}",
    eprint = "2111.03071",
    archivePrefix = "arXiv",
    primaryClass = "cond-mat.stat-mech",
    doi = "10.21468/SciPostPhys.12.6.190",
    journal = "SciPost Phys.",
    volume = "12",
    number = "6",
    pages = "190",
    year = "2022"
}

@article{Grinstein:2018xwp,
    author = "Grinstein, Benjam{\'\i}n and Pal, Sridip",
    title = "{Existence and construction of Galilean invariant $z\neq2$ theories}",
    eprint = "1803.03676",
    archivePrefix = "arXiv",
    primaryClass = "hep-th",
    doi = "10.1103/PhysRevD.97.125006",
    journal = "Phys. Rev. D",
    volume = "97",
    number = "12",
    pages = "125006",
    year = "2018"
}

@article{Das:2009fb,
    author = "Das, Sumit R. and Murthy, Ganpathy",
    title = "{Compact z=2 Electrodynamics in 2+1 dimensions: Confinement with gapless modes}",
    eprint = "0909.3064",
    archivePrefix = "arXiv",
    primaryClass = "hep-th",
    reportNumber = "UK-09-10",
    doi = "10.1103/PhysRevLett.104.181601",
    journal = "Phys. Rev. Lett.",
    volume = "104",
    pages = "181601",
    year = "2010"
}

@article{Dhar:2009am,
    author = "Dhar, Avinash and Mandal, Gautam and Nag, Partha",
    title = "{Renormalization group flows in a Lifshitz-like four fermi model}",
    eprint = "0911.5316",
    archivePrefix = "arXiv",
    primaryClass = "hep-th",
    reportNumber = "TIFR-TH-09-42",
    doi = "10.1103/PhysRevD.81.085005",
    journal = "Phys. Rev. D",
    volume = "81",
    pages = "085005",
    year = "2010"
}

@article{Arav:2016akx,
    author = "Arav, Igal and Oz, Yaron and Raviv-Moshe, Avia",
    title = "{Lifshitz Anomalies, Ward Identities and Split Dimensional Regularization}",
    eprint = "1612.03500",
    archivePrefix = "arXiv",
    primaryClass = "hep-th",
    doi = "10.1007/JHEP03(2017)088",
    journal = "JHEP",
    volume = "03",
    pages = "088",
    year = "2017"
}

@article{Hofman:2011zj,
    author = "Hofman, Diego M. and Strominger, Andrew",
    title = "{Chiral Scale and Conformal Invariance in 2D Quantum Field Theory}",
    eprint = "1107.2917",
    archivePrefix = "arXiv",
    primaryClass = "hep-th",
    doi = "10.1103/PhysRevLett.107.161601",
    journal = "Phys. Rev. Lett.",
    volume = "107",
    pages = "161601",
    year = "2011"
}

@article{Polchinski:1987dy,
    author = "Polchinski, Joseph",
    title = "{Scale and Conformal Invariance in Quantum Field Theory}",
    reportNumber = "UTTG-22-87",
    doi = "10.1016/0550-3213(88)90179-4",
    journal = "Nucl. Phys. B",
    volume = "303",
    pages = "226--236",
    year = "1988"
}

@article{Henkel:1997zz,
    author = "Henkel, Malte",
    title = "{Local Scale Invariance and Strongly Anisotropic Equilibrium Critical Systems}",
    eprint = "cond-mat/9610174",
    archivePrefix = "arXiv",
    doi = "10.1103/PhysRevLett.78.1940",
    journal = "Phys. Rev. Lett.",
    volume = "78",
    pages = "1940--1943",
    year = "1997"
}

@article{Dijkgraaf:1996iy,
    author = "Dijkgraaf, Robbert",
    title = "{Chiral deformations of conformal field theories}",
    eprint = "hep-th/9609022",
    archivePrefix = "arXiv",
    doi = "10.1016/S0550-3213(97)00153-3",
    journal = "Nucl. Phys. B",
    volume = "493",
    pages = "588--612",
    year = "1997"
}

@article{Nahum:2025bdg,
    author = "Nahum, Adam",
    title = "{Continuous symmetry breaking in 1D spin chains and 1+1D field theory}",
    eprint = "2506.21540",
    archivePrefix = "arXiv",
    primaryClass = "cond-mat.stat-mech",
    month = "6",
    year = "2025"
}

@article{Baiguera:2023fus,
    author = "Baiguera, Stefano",
    title = "{Aspects of non-relativistic quantum field theories}",
    eprint = "2311.00027",
    archivePrefix = "arXiv",
    primaryClass = "hep-th",
    doi = "10.1140/epjc/s10052-024-12630-y",
    journal = "Eur. Phys. J. C",
    volume = "84",
    number = "3",
    pages = "268",
    year = "2024"
}

@article{Son:2007ja,
    author = "Son, D. T.",
    title = "{Quantum critical point in graphene approached in the limit of infinitely strong Coulomb interaction}",
    eprint = "cond-mat/0701501",
    archivePrefix = "arXiv",
    reportNumber = "INT-PUB-07-07",
    doi = "10.1103/PhysRevB.75.235423",
    journal = "Phys. Rev. B",
    volume = "75",
    number = "23",
    pages = "235423",
    year = "2007"
}

@article{Nakayama:2013ssa,
    author = "Nakayama, Yu",
    title = "{Vector Beta function}",
    eprint = "1310.0574",
    archivePrefix = "arXiv",
    primaryClass = "hep-th",
    reportNumber = "IPMU13-0186",
    doi = "10.1142/S0217751X13501662",
    journal = "Int. J. Mod. Phys. A",
    volume = "28",
    pages = "1350166",
    year = "2013"
}

@article{Korovin:2013bua,
    author = "Korovin, Yegor and Skenderis, Kostas and Taylor, Marika",
    title = "{Lifshitz as a deformation of Anti-de Sitter}",
    eprint = "1304.7776",
    archivePrefix = "arXiv",
    primaryClass = "hep-th",
    doi = "10.1007/JHEP08(2013)026",
    journal = "JHEP",
    volume = "08",
    pages = "026",
    year = "2013"
}

@article{Boisvert:2025hex,
    author = "Boisvert, Mathieu and Fadda, Shehab Hossam and Kulp, Justin and Yazdi, Ramtin M.",
    title = {{Revisiting Schr{\"o}dinger CFTs: Factorization, Massless Particles, and a Path to the Bootstrap}},
    eprint = "2510.26872",
    archivePrefix = "arXiv",
    primaryClass = "hep-th",
    month = "10",
    year = "2025"
}

@article{DAllen_2001,
doi = {10.1088/0305-4470/34/21/101},
url = {https://doi.org/10.1088/0305-4470/34/21/101},
year = {2001},
month = {jun},
publisher = {},
volume = {34},
number = {21},
pages = {L305},
author = {D Allen and P Azaria and P Lecheminant},
title = {A two-leg quantum Ising ladder: a bosonization study of
the ANNNI model},
journal = {Journal of Physics A: Mathematical and General}
}

@article{SELKE1988213,
title = {The ANNNI model — Theoretical analysis and experimental application},
journal = {Physics Reports},
volume = {170},
number = {4},
pages = {213-264},
year = {1988},
issn = {0370-1573},
doi = {https://doi.org/10.1016/0370-1573(88)90140-8},
url = {https://www.sciencedirect.com/science/article/pii/0370157388901408},
author = {Walter Selke}
}

@article{Dimer,
  title = {Superconductivity and the Quantum Hard-Core Dimer Gas},
  author = {Rokhsar, Daniel S. and Kivelson, Steven A.},
  journal = {Phys. Rev. Lett.},
  volume = {61},
  issue = {20},
  pages = {2376--2379},
  numpages = {0},
  year = {1988},
  month = {Nov},
  publisher = {American Physical Society},
  doi = {10.1103/PhysRevLett.61.2376},
  url = {https://link.aps.org/doi/10.1103/PhysRevLett.61.2376}
}

@article{Dimer2,
  title = {Dynamical stability of the quantum Lifshitz theory in 2+1 dimensions},
  author = {Hsu, Benjamin and Fradkin, Eduardo},
  journal = {Phys. Rev. B},
  volume = {87},
  issue = {8},
  pages = {085102},
  numpages = {16},
  year = {2013},
  month = {Feb},
  publisher = {American Physical Society},
  doi = {10.1103/PhysRevB.87.085102},
  url = {https://link.aps.org/doi/10.1103/PhysRevB.87.085102}
}

@article{BoseHubbard,
  title = {Boson localization and the superfluid-insulator transition},
  author = {Fisher, Matthew P. A. and Weichman, Peter B. and Grinstein, G. and Fisher, Daniel S.},
  journal = {Phys. Rev. B},
  volume = {40},
  issue = {1},
  pages = {546--570},
  numpages = {0},
  year = {1989},
  month = {Jul},
  publisher = {American Physical Society},
  doi = {10.1103/PhysRevB.40.546},
  url = {https://link.aps.org/doi/10.1103/PhysRevB.40.546}
}

@article{FermiUnitarity,
  title = {Universal Thermodynamics of Degenerate Quantum Gases in the Unitarity Limit},
  author = {Ho, Tin-Lun},
  journal = {Phys. Rev. Lett.},
  volume = {92},
  issue = {9},
  pages = {090402},
  numpages = {4},
  year = {2004},
  month = {Mar},
  publisher = {American Physical Society},
  doi = {10.1103/PhysRevLett.92.090402},
  url = {https://link.aps.org/doi/10.1103/PhysRevLett.92.090402}
}

@article{FermiSon,
  title = {Vanishing Bulk Viscosities and Conformal Invariance of the Unitary Fermi Gas},
  author = {Son, D. T.},
  journal = {Phys. Rev. Lett.},
  volume = {98},
  issue = {2},
  pages = {020604},
  numpages = {4},
  year = {2007},
  month = {Jan},
  publisher = {American Physical Society},
  doi = {10.1103/PhysRevLett.98.020604},
  url = {https://link.aps.org/doi/10.1103/PhysRevLett.98.020604}
}

@article{NRSon,
  title = {Nonrelativistic conformal field theories},
  author = {Nishida, Yusuke and Son, Dam T.},
  journal = {Phys. Rev. D},
  volume = {76},
  issue = {8},
  pages = {086004},
  numpages = {14},
  year = {2007},
  month = {Oct},
  publisher = {American Physical Society},
  doi = {10.1103/PhysRevD.76.086004},
  url = {https://link.aps.org/doi/10.1103/PhysRevD.76.086004}
}

@article{Cappelli:1986hf,
    author = "Cappelli, Andrea and Itzykson, C. and Zuber, J. -B.",
    title = "{Modular invariant partition functions in two dimensions}",
    reportNumber = "SACLAY-PHT-86-122",
    doi = "10.1016/0550-3213(87)90155-6",
    journal = "Nucl. Phys. B",
    volume = "280",
    pages = "445--465",
    year = "1987"
}

@article{Baxter:1989vv,
    author = "Baxter, R. J.",
    title = "{Superintegrable chiral Potts model: Thermodynamic properties, an inverse model and a simple associated Hamiltonian}",
    doi = "10.1007/BF01023632",
    journal = "J. Statist. Phys.",
    volume = "57",
    pages = "1--39",
    year = "1989"
}

@article{Antunes:2025huk,
    author = "Antunes, Ant{\'o}nio and Rong, Junchen",
    title = "{Irrational CFTs from coupled anyon chains with non-invertible symmetries?}",
    eprint = "2507.14280",
    archivePrefix = "arXiv",
    primaryClass = "hep-th",
    month = "7",
    year = "2025"
}

@article{Antunes:2025erb,
    author = "Antunes, Ant{\'o}nio and Suchel, No{\'e}",
    title = "{Taxonomy of coupled minimal models from finite groups}",
    eprint = "2512.23664",
    archivePrefix = "arXiv",
    primaryClass = "hep-th",
    month = "12",
    year = "2025"
}

@article{Komargodski:2016auf,
    author = "Komargodski, Zohar and Simmons-Duffin, David",
    title = "{The Random-Bond Ising Model in 2.01 and 3 Dimensions}",
    eprint = "1603.04444",
    archivePrefix = "arXiv",
    primaryClass = "hep-th",
    doi = "10.1088/1751-8121/aa6087",
    journal = "J. Phys. A",
    volume = "50",
    number = "15",
    pages = "154001",
    year = "2017"
}

@article{Wang_2022,
   title={Universal finite-size amplitude and anomalous entangment entropy of $z=2$ quantum Lifshitz criticalities in topological chains},
   volume={12},
   ISSN={2542-4653},
   url={http://dx.doi.org/10.21468/SciPostPhys.12.4.134},
   DOI={10.21468/scipostphys.12.4.134},
   number={4},
   journal={SciPost Physics},
   publisher={Stichting SciPost},
   author={Wang, Ke and Sedrakyan, Tigran},
   year={2022},
   month=apr }

@article{Osborn:2020cnf,
    author = "Osborn, Hugh and Stergiou, Andreas",
    title = "{Heavy handed quest for fixed points in multiple coupling scalar theories in the $\epsilon$ expansion}",
    eprint = "2010.15915",
    archivePrefix = "arXiv",
    primaryClass = "hep-th",
    reportNumber = "LA-UR-20-27569",
    doi = "10.1007/JHEP04(2021)128",
    journal = "JHEP",
    volume = "04",
    pages = "128",
    year = "2021"
}

@article{Osborn:2017ucf,
    author = "Osborn, Hugh and Stergiou, Andreas",
    title = "{Seeking fixed points in multiple coupling scalar theories in the $\epsilon$ expansion}",
    eprint = "1707.06165",
    archivePrefix = "arXiv",
    primaryClass = "hep-th",
    reportNumber = "DAMTP-2017-30, CERN-TH-2017-149",
    doi = "10.1007/JHEP05(2018)051",
    journal = "JHEP",
    volume = "05",
    pages = "051",
    year = "2018"
}

@article{Zamolodchikov:1987ti,
    author = "Zamolodchikov, A. B.",
    title = "{Renormalization Group and Perturbation Theory Near Fixed Points in Two-Dimensional Field Theory}",
    journal = "Sov. J. Nucl. Phys.",
    volume = "46",
    pages = "1090",
    year = "1987"
}

@article{Antunes:2022vtb,
    author = "Antunes, Ant{\'o}nio and Behan, Connor",
    title = "{Coupled Minimal Conformal Field Theory Models Revisited}",
    eprint = "2211.16503",
    archivePrefix = "arXiv",
    primaryClass = "hep-th",
    reportNumber = "DESY-22-191",
    doi = "10.1103/PhysRevLett.130.071602",
    journal = "Phys. Rev. Lett.",
    volume = "130",
    number = "7",
    pages = "071602",
    year = "2023"
}

@article{Antunes:2024mfb,
    author = "Antunes, Ant{\'o}nio and Behan, Connor",
    title = "{Coupled minimal models revisited II: Constraints from permutation symmetry}",
    eprint = "2412.21107",
    archivePrefix = "arXiv",
    primaryClass = "hep-th",
    doi = "10.21468/SciPostPhys.18.4.132",
    journal = "SciPost Phys.",
    volume = "18",
    number = "4",
    pages = "132",
    year = "2025"
}

@article{Kousvos:2024dlz,
    author = "Kousvos, Stefanos R. and Piazza, Alessandro and Vichi, Alessandro",
    title = "{Exploring replica-Potts CFTs in two dimensions}",
    eprint = "2405.19416",
    archivePrefix = "arXiv",
    primaryClass = "hep-th",
    doi = "10.1007/JHEP11(2024)030",
    journal = "JHEP",
    volume = "11",
    pages = "030",
    year = "2024"
}

@article{Chang:2018iay,
    author = "Chang, Chi-Ming and Lin, Ying-Hsuan and Shao, Shu-Heng and Wang, Yifan and Yin, Xi",
    title = "{Topological Defect Lines and Renormalization Group Flows in Two Dimensions}",
    eprint = "1802.04445",
    archivePrefix = "arXiv",
    primaryClass = "hep-th",
    reportNumber = "CALT-TH-2017-067, CALT-TH 2017-067, PUPT-2546",
    doi = "10.1007/JHEP01(2019)026",
    journal = "JHEP",
    volume = "01",
    pages = "026",
    year = "2019"
}

@article{Rychkov:2018vya,
    author = "Rychkov, Slava and Stergiou, Andreas",
    title = "{General Properties of Multiscalar RG Flows in $d=4-\varepsilon$}",
    eprint = "1810.10541",
    archivePrefix = "arXiv",
    primaryClass = "hep-th",
    reportNumber = "CERN-TH-2018-225",
    doi = "10.21468/SciPostPhys.6.1.008",
    journal = "SciPost Phys.",
    volume = "6",
    number = "1",
    pages = "008",
    year = "2019"
}

@article{Hogervorst:2020gtc,
    author = "Hogervorst, Matthijs and Toldo, Chiara",
    title = "{Bounds on multiscalar CFTs in the $\epsilon$ expansion}",
    eprint = "2010.16222",
    archivePrefix = "arXiv",
    primaryClass = "hep-th",
    doi = "10.1007/JHEP04(2021)068",
    journal = "JHEP",
    volume = "04",
    pages = "068",
    year = "2021"
}

@article{Rong:2023xhz,
    author = "Rong, Junchen and Rychkov, Slava",
    title = "{Classifying irreducible fixed points of five scalar fields in perturbation theory}",
    eprint = "2306.09419",
    archivePrefix = "arXiv",
    primaryClass = "hep-th",
    doi = "10.21468/SciPostPhys.16.2.040",
    journal = "SciPost Phys.",
    volume = "16",
    number = "2",
    pages = "040",
    year = "2024"
}

@article{MerminWagner,
  title = {Absence of Ferromagnetism or Antiferromagnetism in One- or Two-Dimensional Isotropic Heisenberg Models},
  author = {Mermin, N. D. and Wagner, H.},
  journal = {Phys. Rev. Lett.},
  volume = {17},
  issue = {22},
  pages = {1133--1136},
  numpages = {0},
  year = {1966},
  month = {Nov},
  publisher = {American Physical Society},
  doi = {10.1103/PhysRevLett.17.1133},
  url = {https://link.aps.org/doi/10.1103/PhysRevLett.17.1133}
}

@article{Coleman:1973ci,
    author = "Coleman, Sidney R.",
    title = "{There are no Goldstone bosons in two-dimensions}",
    doi = "10.1007/BF01646487",
    journal = "Commun. Math. Phys.",
    volume = "31",
    pages = "259--264",
    year = "1973"
}

@article{Nakayama:2013is,
    author = "Nakayama, Yu",
    title = "{Scale invariance vs conformal invariance}",
    eprint = "1302.0884",
    archivePrefix = "arXiv",
    primaryClass = "hep-th",
    reportNumber = "CALT-68-2910",
    doi = "10.1016/j.physrep.2014.12.003",
    journal = "Phys. Rept.",
    volume = "569",
    pages = "1--93",
    year = "2015"
}

@article{Dymarsky:2013pqa,
    author = "Dymarsky, Anatoly and Komargodski, Zohar and Schwimmer, Adam and Theisen, Stefan",
    title = "{On Scale and Conformal Invariance in Four Dimensions}",
    eprint = "1309.2921",
    archivePrefix = "arXiv",
    primaryClass = "hep-th",
    doi = "10.1007/JHEP10(2015)171",
    journal = "JHEP",
    volume = "10",
    pages = "171",
    year = "2015"
}

@article{Gimenez-Grau:2023lpz,
    author = "Gimenez-Grau, Aleix and Nakayama, Yu and Rychkov, Slava",
    title = "{Scale without conformal invariance in dipolar ferromagnets}",
    eprint = "2309.02514",
    archivePrefix = "arXiv",
    primaryClass = "hep-th",
    reportNumber = "YITP-23-107",
    doi = "10.1103/PhysRevB.110.024421",
    journal = "Phys. Rev. B",
    volume = "110",
    number = "2",
    pages = "024421",
    year = "2024"
}

@article{PhysRevLett.28.240,
  title = {Critical Exponents in 3.99 Dimensions},
  author = {Wilson, Kenneth G. and Fisher, Michael E.},
  journal = {Phys. Rev. Lett.},
  volume = {28},
  issue = {4},
  pages = {240--243},
  numpages = {0},
  year = {1972},
  month = {Jan},
  publisher = {American Physical Society},
  doi = {10.1103/PhysRevLett.28.240},
  url = {https://link.aps.org/doi/10.1103/PhysRevLett.28.240}
}

@article{Behan:2017rca,
    author = "Behan, Connor",
    title = "{Unitary subsector of generalized minimal models}",
    eprint = "1712.06622",
    archivePrefix = "arXiv",
    primaryClass = "hep-th",
    reportNumber = "YITP-SB-17-54, YITP-SB-17-54",
    doi = "10.1103/PhysRevD.97.094020",
    journal = "Phys. Rev. D",
    volume = "97",
    number = "9",
    pages = "094020",
    year = "2018"
}

@article{Ribault:2018jdv,
    author = "Ribault, Sylvain",
    title = "{On 2d CFTs that interpolate between minimal models}",
    eprint = "1809.03722",
    archivePrefix = "arXiv",
    primaryClass = "hep-th",
    doi = "10.21468/SciPostPhys.6.6.075",
    journal = "SciPost Phys.",
    volume = "6",
    number = "6",
    pages = "075",
    year = "2019"
}

@article{Ribault:2019qrz,
    author = "Ribault, Sylvain",
    title = "{The non-rational limit of D-series minimal models}",
    eprint = "1909.10784",
    archivePrefix = "arXiv",
    primaryClass = "hep-th",
    doi = "10.21468/SciPostPhysCore.3.1.002",
    journal = "SciPost Phys. Core",
    volume = "3",
    pages = "002",
    year = "2020"
}

@article{Hellerman:2021qzz,
    author = "Hellerman, Simeon and Orlando, Domenico and Pellizzani, Vito and Reffert, Susanne and Swanson, Ian",
    title = "{Nonrelativistic CFTs at large charge: Casimir energy and logarithmic enhancements}",
    eprint = "2111.12094",
    archivePrefix = "arXiv",
    primaryClass = "hep-th",
    doi = "10.1007/JHEP05(2022)135",
    journal = "JHEP",
    volume = "05",
    pages = "135",
    year = "2022"
}

@article{Bargmann:1954gh,
    author = "Bargmann, V.",
    title = "{On Unitary ray representations of continuous groups}",
    doi = "10.2307/1969831",
    journal = "Annals Math.",
    volume = "59",
    pages = "1--46",
    year = "1954"
}

@article{Hellerman:2023myh,
    author = "Hellerman, Simeon and Krichevskiy, Daniil and Orlando, Domenico and Pellizzani, Vito and Reffert, Susanne and Swanson, Ian",
    title = "{The unitary Fermi gas at large charge and large N}",
    eprint = "2311.14793",
    archivePrefix = "arXiv",
    primaryClass = "hep-th",
    doi = "10.1007/JHEP05(2024)323",
    journal = "JHEP",
    volume = "05",
    pages = "323",
    year = "2024"
}

@article{Dotsenko:1984nm,
	author = "Dotsenko, V. S. and Fateev, V. A.",
	editor = "Khalatnikov, I. M. and Mineev, V. P.",
	title = "{Conformal Algebra and Multipoint Correlation Functions in Two-Dimensional Statistical Models}",
	reportNumber = "NORDITA-84/8",
	doi = "10.1016/0550-3213(84)90269-4",
	journal = "Nucl. Phys. B",
	volume = "240",
	pages = "312",
	year = "1984"
}

@article{Kravchuk:2025evf,
    author = "Kravchuk, Petr and Radcliffe, Alex",
    title = "{Monodromy Pinning Defects in the Critical $\mathrm{O}(2N)$ Model}",
    eprint = "2510.02281",
    archivePrefix = "arXiv",
    primaryClass = "hep-th",
    month = "10",
    year = "2025"
}

@article{Drukker:2025dfm,
    author = "Drukker, Nadav and Kong, Ziwen and Kravchuk, Petr",
    title = "{Nonlinearly Realised Defect Symmetries and Anomalies}",
    eprint = "2512.15913",
    archivePrefix = "arXiv",
    primaryClass = "hep-th",
    month = "12",
    year = "2025"
}

@article{Dotsenko:1984ad,
    author = "Dotsenko, V. S. and Fateev, V. A.",
    title = "{Four Point Correlation Functions and the Operator Algebra in the Two-Dimensional Conformal Invariant Theories with the Central Charge c \ensuremath{<} 1}",
    reportNumber = "NORDITA-84/22",
    doi = "10.1016/S0550-3213(85)80004-3",
    journal = "Nucl. Phys. B",
    volume = "251",
    pages = "691--734",
    year = "1985"
}

@article{Zamolodchikov:1986gt,
    author = "Zamolodchikov, A. B.",
    title = "{Irreversibility of the Flux of the Renormalization Group in a 2D Field Theory}",
    journal = "JETP Lett.",
    volume = "43",
    pages = "730--732",
    year = "1986"
}

@article{Cardy:1988tj,
    author = "Cardy, John L.",
    title = "{The Central Charge and Universal Combinations of Amplitudes in Two-dimensional Theories Away From Criticality}",
    reportNumber = "UCSB-TH-89-1988",
    doi = "10.1103/PhysRevLett.60.2709",
    journal = "Phys. Rev. Lett.",
    volume = "60",
    pages = "2709",
    year = "1988"
}

@article{Drukker:2022pxk,
    author = "Drukker, Nadav and Kong, Ziwen and Sakkas, Georgios",
    title = "{Broken Global Symmetries and Defect Conformal Manifolds}",
    eprint = "2203.17157",
    archivePrefix = "arXiv",
    primaryClass = "hep-th",
    doi = "10.1103/PhysRevLett.129.201603",
    journal = "Phys. Rev. Lett.",
    volume = "129",
    number = "20",
    pages = "201603",
    year = "2022"
}

@article{Gabai:2025zcs,
    author = "Gabai, Barak and Sever, Amit and Zhong, De-liang",
    title = "{Universal constraints for conformal line defects}",
    eprint = "2501.06900",
    archivePrefix = "arXiv",
    primaryClass = "hep-th",
    doi = "10.1103/gsfg-wrps",
    journal = "Phys. Rev. D",
    volume = "112",
    number = "6",
    pages = "065004",
    year = "2025"
}

@article{Girault:2025kzt,
    author = "Girault, Bastien and Paulos, Miguel F. and van Vliet, Philine",
    title = "{Consequences of symmetry-breaking on conformal defect data}",
    eprint = "2509.26561",
    archivePrefix = "arXiv",
    primaryClass = "hep-th",
    month = "9",
    year = "2025"
}

@article{Jack:1990eb,
    author = "Jack, I. and Osborn, H.",
    title = "{Analogs for the $c$ Theorem for Four-dimensional Renormalizable Field Theories}",
    reportNumber = "DAMTP-90-02",
    doi = "10.1016/0550-3213(90)90584-Z",
    journal = "Nucl. Phys. B",
    volume = "343",
    pages = "647--688",
    year = "1990"
}
\bibliographystyle{utphys}

\end{document}